\documentclass[aps,prd,twocolumn,floatfix,showpacs]{revtex4} 

\usepackage{graphics}
\usepackage{graphicx}

\usepackage{latexsym}
\usepackage[cp866]{inputenc}
\usepackage[english]{babel}

\input epsf

\begin{document}

\title{\boldmath $\eta(1475)$ and $f_1(1420)$ resonances in
$\gamma\gamma^*$ collisions and
$J/\psi\to\gamma(\rho\rho,\,\gamma\rho^0,\,\gamma\phi)$ decays}
\author{N.N. Achasov\footnote{E-mail: achasov@math.nsc.ru}
and G.N. Shestakov\footnote{E-mail: shestako@math.nsc.ru}
}\affiliation{Laboratory of  Theoretical Physics, S.L. Sobolev
Institute for Mathematics, 630090, Novosibirsk, Russia}


\begin{abstract}
The available data on the $Q^2$ dependence of the $\gamma\gamma^*(
Q^2)\to K\bar K\pi$ reaction cross section in the energy region
1.35--1.55 GeV is explained by the $\eta(1475)$ resonance production
in contrast to their conventional interpretation with the use of the
$f_1(1420)$ resonance. It resolves theoretically the contradiction
between the suppression of the $\eta(1475)\to\gamma\gamma$ decay
width and the strong couplings of the $\eta(1475)$ to the $\rho
\rho$, $\omega\omega$, and $\gamma\rho^0$ channels. The experimental
check of our explanation requires definition of the spin-parity of
the resonance contributions, $R$, in $\gamma\gamma^*(Q^2)\to R\to
K\bar K\pi$ and in $J/\psi\to\gamma R\to\gamma\gamma(\rho^0, \phi)$.
This will help to solve difficulties accumulated in understanding
properties of the $\eta(1475)$ state and its nearest partners.
\end{abstract}


\maketitle

\section{INTRODUCTION}

The family of the pseudoscalar states $\pi(1300)$, $\eta(1295)$,
$\eta(1405)$, $\eta(1475)$, and $K(1460)$ \cite{PDG10} has very
mysterious properties; see for review \cite{KW,Ac1,GN,PDG02,AT,
PDG04,MCU,PDG06,KZ,AM}. The $\eta(1475)$ can be also confused with
the $f_1(1420)$ \cite{PDG10} in the $K\bar K\pi$ decay channel,
without carrying out a partial wave analysis.

Recall that in 2004 the long-known state E$/\iota(1440)/\eta(1440)$
was officially split into two components \cite{PDG04,PDG06},
$\eta(1405)$ and $\eta(1475)$, decaying mainly into $a_0(980)\pi$
and $K^*(892)\bar K$, respectively \cite{PDG10,MCU,AM}. The
splitting of the $\eta(1440)$ significantly complicates the
classification of the above-mentioned pseudoscalar states
\cite{PDG10,MCU,PDG06,KZ,AM}. An ``odd'' member of this family with
putative glueball properties is often associated with the
$\eta(1405)$ state \cite{PDG10,MCU,AM}. Notice that this state has
not been seen in $\gamma\gamma$ or in $\gamma\gamma^*$ collisions
even in its main decay channel into $\eta\pi\pi$
\cite{PDG10,MCU,AM,Ac,Ach}. The available data on the
$J/\psi\to\gamma\eta(1440)\to\gamma\gamma\rho^0$ and
$J/\psi\to\gamma\eta(1440)\to\gamma\rho^0\rho^0$ decays are quoted
in the later Particle Data Group (PDG) reviews
\cite{PDG10,PDG04,PDG06}, in the $J/\psi(1S)$ section, as the data
for the unresolved $\eta(1405/1475)$ state. At the same time, the
even data on $J/\psi\to\gamma\eta(1440)\to\gamma\gamma\rho^0$ are
assigned to the $\eta(1405)$ state in the $\eta(1405)$ section
\cite{PDG10}. To clarify the existing uncertain situation, we
attribute the data on $J/\psi\to\gamma\eta(1440)\to\gamma
(\gamma\rho^0, \rho^0\rho^0)$ decays to the $\eta(1475)$. (If we
were to attribute these decays to $\eta(1405)$, then our scenario
would not change.) Experimental verification of our scenario will
automatically resolve the question of the role of the $\eta(1405)$
in the problem under discussion. We emphasize that, if the L3
Collaboration is right when they assert that the $\eta(1405)$ signal
is absent in $\gamma\gamma$ and $\gamma\gamma^*$ collisions
\cite{Ac,Ach}, then most likely the data on the $\gamma\rho^0$ decay
cannot be attributed to the $\eta(1405)$.

In this paper we concentrate on the manifestations of the
$\eta(1475)$ and $f_1(1420)$ resonances in $\gamma\gamma$ and
$\gamma\gamma^*(Q^2)$ collisions (where $Q^2$ is a photon
virtuality) and in $J/\psi\to\gamma\gamma(\rho^0,\phi)$ decays.
Section II describes the history of the search for the $\eta(1475)$
meson production in $\gamma\gamma$ collisions. In Sec. III, the
paradox related to the suppression of the $\eta(1475)\to\gamma
\gamma$ decay width and the strong couplings of the $\eta(1475)$ to
the $\rho\rho$, $\omega \omega$, and $\gamma\rho^0$ channels is
presented. In Sec. IV, a scenario resolving the problem with the
$\eta(1475)\to\gamma\gamma$ decay is considered, and on its basis we
obtain the alternative explanation of the existing data on the $Q^2$
dependence of the $\gamma\gamma^* (Q^2)\to K\bar K\pi$ reaction
cross section in the region of the $\eta(1475)$ and $f_1(1420)$
resonances. The relations between the data on the $f_1(1420)\to
\gamma(\rho^0,\phi)$ and $f_1(1420)\to\gamma\gamma^*$ decays are
discussed in Sec. V. The measurements required to solve accumulated
difficulties in understanding properties of the $\eta(1475)$ state
and to check our explanation are briefly listed in Sec. VI. Several
technical details are relegated to the Appendix.

\section{\boldmath THE $\eta(1475)\to\gamma\gamma$ DECAY}

\begin{table} 
\caption{Information about the $\eta(1475)\to\gamma\gamma$ decay.}
\vspace{0.05cm} 
\begin{tabular}{|c|c|c|}
\hline Experiment & Ref. & $\ \Gamma(\eta(1475)\to\gamma\gamma)\ $
\\ &  & $\ \times B(\eta(1475)\to K\bar K\pi)$ (keV)\ \\ \hline
MARK II,       1983 & \cite{Je}  &  $<8$                      \\
TASSO,         1985 & \cite{Al}  &  $<2.2$                    \\
TPC/2$\gamma$, 1986 & \cite{Ai}  &  $<1.6$                    \\
CELLO,         1989 & \cite{Be}  &  $<1.2$                    \\
L3,            2001 & \cite{Ac}  &  $0.212\pm0.050\pm0.023$   \\
CLEO II,       2005 & \cite{Ah}  &  $<0.089$ (90\% C.L.)      \\
L3,            2007 & \cite{Ach} &  $0.23\pm0.05\pm0.05$      \\
\hline Experiment & Ref. & $\Gamma(\eta(1475)\to\gamma\gamma)\ $ \\
&  & $\ \times B(\eta(1475)\to\eta\pi\pi)$ (keV)\
\\ \hline
\ Crystal Ball,  1987\  & \cite{An}  &  $<0.3$                \\
L3,            2001 & \cite{Ac}  &  $<0.095$ (95\% C.L.)      \\
\hline\end{tabular}\end{table}

The history of the search for the $\eta(1475)$ meson production in
$\gamma\gamma$ collisions in the $K\bar K\pi$ and $\eta\pi\pi$ final
states is represented in Table I \cite{Ac,Ach,Je,Al,Ai,Be,Ah,An}.
The first successful experiment was performed by the L3
Collaboration \cite{Ac}, in which $37\pm9$
$\gamma\gamma\to\eta(1475)\to K^0_SK^\pm\pi^\mp$ events were
selected in the region of the resonance peak with the mass
$M=1481\pm12$ MeV. Then the CLEO II Collaboration \cite{Ah}
investigated the reaction $\gamma\gamma\to K^0_SK^\pm \pi^\mp$ with
a data sample that exceeds the L3 statistics by a factor of 5 and
did not find any signal in this region. As a result, they obtained
the strongest upper limit on the product of the
$\eta(1475)\to\gamma\gamma$ decay width, $\Gamma(\eta(1475)\to
\gamma \gamma)$, and the $\eta(1475)\to K\bar K\pi$ decay branching
ratio, $B(\eta(1475)\to K\bar K\pi)$; see Table I. Later on, the L3
Collaboration treated all collected statistics (which is 50\% higher
than that used previously \cite{Ac})) and confirmed the result of
their own experiment, $\Gamma(\eta(1475)\to\gamma\gamma)
B(\eta(1475)\to K\bar K\pi)=0.23\pm0.05 \pm0.05$ keV \cite{Ach,FN1}.
As noted in Ref. \cite{Ach}, if the world average width of
$\eta(1475)$ is used, the CLEO II upper limit increases from 0.089
keV to 0.140 keV that is consistent with the L3 result within two
errors.

\section{\boldmath PARADOX OF THE
$\eta(1475)\to\rho^0\rho^0$, $\eta(1475)\to\gamma\rho^0$, and
$\eta(1475)\to\gamma\gamma$ DECAYS}

The $J/\psi$ meson radiative decays, such as $J/\psi\to\gamma K\bar
K\pi$, $J/\psi\to\gamma\rho\rho$, $J/\psi\to\gamma\omega\omega$, and
$J/\psi\to\gamma\gamma\rho^0$, are the very important tool of the
$\eta(1475) $ meson investigation \cite{PDG10,KW,Ac1,MCU,Bur,Hi,
Pe,Wer,Ri,AS1,We,Ba2,Ba1,Bi2,Bi1,Co,Au,B2,B3,B1}.

In the mid 1980s, we showed \cite{AS1} that pseudoscalar ($J^P=0^-$)
structures discovered by MARK III \cite{Ri,Bur,Wer} in the
$\rho\rho$ and $\omega\omega$ mass spectra near their thresholds in
the $J/\psi\to\gamma\rho\rho$ and $J/\psi\to\gamma\omega\omega$
decays can be explained by decays $\eta(1440)\to\rho\rho$ and
$\eta(1440)\to\omega\omega$ at the resonance tail. This explanation
was supported by subsequent results from MARK III
\cite{KW,We,Ba1,Ba2}, DM2 \cite{Bi1,Bi2}, and BES \cite{B2} on the
$J/\psi\to\gamma\rho\rho$ and $J/\psi\to\gamma\omega\omega$ decays.
In the work \cite{AS1}, we also showed that the strong coupling of
$\eta(1440)$ to $\rho^0\rho^0$ leads within the usual vector
dominance model (VDM) to the large decay widths $\eta(1440)\to\gamma
\gamma$ and $\eta(1440)\to\gamma\rho^0$: $\Gamma(\eta(1440)\to
\rho^0\rho^0\to\gamma\gamma)\approx$ 6.6 keV and $\Gamma(\eta(1440)
\to\gamma\rho^0)\approx$ 1.3 MeV. Note that these values should be
doubled at present because the branching ratio
$B(J/\psi\to\gamma\eta(1405 /1475)\to\gamma\rho^0\rho^0)$\,=\,$
(1.7\pm0.4)\times10^{-3}$ \cite{PDG10} has since increased
approximately 2 times. Such an estimate for the width
$\Gamma(\eta(1440)\to\gamma\gamma)$ is in apparent contradiction
with the results of its direct measurements presented in Table I.
The recent experiments performed by L3 \cite{Ac,Ach} and CLEO II
\cite{Ah} Collaborations essentially sharpened this contradiction
when compared to its first manifestations which were discussed more
than 20 years ago \cite{Hi,Ac1,KW,Ch1,MPP,Bar,Zi,Cl,AS1, AS2}.

Let us consider the data on the $J/\psi\to\gamma\gamma\rho^0$ decay
\cite{PDG10,KW,Hi,Pe,Ri,We,Co,Au,B1} revealing the coupling of the
$\eta(1475)$ to the $\gamma\rho^0$ decay channel. Notice that the
first Crystal Ball and MARK III measurements \cite{Pe,We,KW,Co} did
not clarify the question about the spin-parity of the resonance
enhancement observable in the $\gamma\rho^0$ system near 1.44 GeV.
The $\gamma\rho^0$ angular distributions agree with the $J^P=0^-$
resonance production but spin-parity $J^P=1^+$ was not excluded.
However, not so long ago, the BES Collaboration \cite{B1} obtained
some indirect indication in favor of the $\eta(1440)$ meson
production in the $J/\psi\to\gamma R\to\gamma\gamma\rho^0$ decay. To
determine whether $R$ is more likely to be the $f_1(1420)$ or the
$\eta(1440)$, they used the measurement by the WA102 Collaboration
\cite{Ba},\begin{equation}\label{Lim1}
B(f_1(1420)\to\gamma\rho^0)/B(f_1(1420)\to K\bar K\pi)<0.02,
\end{equation}
together with the PDG result \cite{PDG02},
$$B(J/\psi\to\gamma f_1(1420)\to\gamma K\bar K\pi)=(7.9
\pm1.3)\times10^{-4},$$ and obtained
\begin{equation}\label{Lim2}B(J/\psi\to\gamma f_1(1420)\to\gamma
\gamma\rho^0)<1.7\times10^{-5}\end{equation} (with 95\% C.L.).
Comparing this restriction with their own measurement of
$$B(J/\psi\to\gamma\eta(1440)\to\gamma\gamma\rho^0)=(1.07\pm0.17
\pm0.11)\times10^{-4}$$ (see Table II), they concluded that $R$ in
the $J/\psi\to\gamma R\to\gamma\gamma\rho^0$ channel should be
predominantly $\eta(1440)$. Note that in Sec. V we shall obtain a
stronger restriction for $B(f_1(1420)\to\gamma\rho^0)/B(
f_1(1420)\to K\bar K\pi)$ in comparison with Eq. (\ref{Lim1}), with
the help of the quark model and the data on the $f_1(1420)\to\gamma
\phi$ decay.

The $\eta(1475)\to\gamma\rho^0$ decay width can be estimated from
the relation
\begin{eqnarray}\label{Getagrho}
\Gamma(\eta(1475)\to\gamma\rho^0)=\Gamma^{tot}_{\eta(1475)}
B(\eta(1475)\to K\bar K\pi)\nonumber \\ \times
\frac{B(J/\psi\to\gamma\eta(1475)\to\gamma\gamma\rho^0
)}{B(J/\psi\to\gamma\eta(1475)\to\gamma K\bar K\pi)}\,.\qquad\ \ \
\end{eqnarray}
If we put $\Gamma^{tot}_{\eta(1475)}=85$ MeV \cite{PDG10},
$B(\eta(1475)\to K\bar K\pi)=0.6$ \cite{FN2} and use the last BES
data \cite{B1,B3} for
$B(J/\psi\to\gamma\eta(1475)\to\gamma\gamma\rho^0)$ and
$B(J/\psi\to\gamma\eta(1475)\to\gamma K\bar K\pi)$ (see Table II),
we find that $\Gamma(\eta(1475)\to\gamma\rho^0)\approx3.3$ MeV. If
we use for $B(J/\psi\to\gamma\eta(1475)\to\gamma \gamma \rho^0)$ and
$B(J/\psi\to\gamma\eta(1475)\to\gamma K\bar K\pi)$ the PDG averages
\cite{PDG06,PDG10} (also indicated in Table II), then we have
$\Gamma(\eta(1475)\to\gamma\rho^0)\approx1.4$ MeV. These values
agree with the first results for $\Gamma(\eta
(1440)\to\gamma\rho^0)$ \cite{Hi,Pe,Ri,We,KW,AS1,Ch1,AS2} obtained
over 20 years ago. We accept as a conservative estimate
\begin{equation}\Gamma(\eta(1475)\to\gamma\rho^0)=1\mbox{ MeV}.
\label{EstimateGigr}\end{equation}

\begin{table} 
\caption{Information about the $\eta(1475/1440)$\,$\to$\,$K\bar
K\pi$, $\gamma\rho^0$, $\gamma\omega$, and $\gamma\phi$ decays.}\vspace{0.05cm} 
\begin{tabular}{|c|c|c|}
\hline Experiment & Ref.  & Data \\ \hline BES, 2000 & \cite{B3} &
$B(J/\psi\to\gamma\eta(1440)\to\gamma
K\bar K\pi)$ \\ & & $=(1.66\pm0.10\pm0.58)\times10^{-3}$  \\
BES, 2004 &        \cite{B1}     &
$B(J/\psi\to\gamma\eta(1440)\to\gamma\gamma\rho^0)$ \\
& & $=(1.07\pm0.17\pm0.11)\times10^{-4}$  \\
BES, 2004 &        \cite{B1}     &
$B(J/\psi\to\gamma\eta(1440)\to\gamma\gamma\phi)$ \\
& & $=(0.31\pm0.30)\times10^{-4}$, \\  & & or $<0.82\times10^{-4}$ (95\% C.L.) \\
PDG, &  \cite{PDG06,PDG10}  & $B(J/\psi\to\gamma\eta(1475)\to\gamma
K\bar K\pi)$ \\ 2006, 2010 & & $=(2.8\pm0.6)\times10^{-3}$  \\
PDG, &  \cite{PDG06,PDG10}  &
$B(J/\psi\to\gamma\eta(1475)\to\gamma\gamma\rho^0)$ \\
2006, 2010 & & $=(0.78\pm0.2)\times10^{-4}$  \\
MARK III, &   \,\cite{Ri,KW}  &
$B(J/\psi\to\gamma\eta(1440)\to\gamma\gamma\omega)$ \\
1985 & & $<2.3\times10^{-4}$ (90\% C.L.) \\
\hline \end{tabular}\end{table}

To estimate the width of the $\eta(1475)\to\gamma\gamma$ decay
caused by the transitions $\eta(1475)\to\gamma V$ (where $V=\rho^0$,
$\omega$, and $\phi$) we apply VDM and $SU(3)$ symmetry, together
with the nonet symmetry assumption for the $V$ meson interactions
and the ideal $\omega-\phi$ mixing. Hereafter, for short, the
$\eta(1475)$ will be denoted in the indices by $\iota$. Thus, for
the coupling constants $g_{\iota\gamma\gamma}$ and
$g_{\iota\gamma\rho}$, one can write the following relation
\cite{AS1,Ch1}:
\begin{equation}g_{\iota\gamma\gamma}=\frac{e}{f_\rho}
g_{\iota\gamma\rho}\left(1+\frac{1}{9}+\frac{2}{9}H(x)\right)
\,,\label{VDM1}\end{equation} where the three terms correspond to a
transition via $\gamma\rho$, $\gamma\omega$ and $\gamma\phi$,
respectively, $f^2_\rho/(4\pi)=\alpha^2m_\rho/[3\Gamma(\rho^0\to
e^+e^-)]=1.96$ \cite{PDG10}, $H(x)=(1-2x)/(1+x)$, the parameter
$x=r\tan\theta_\iota/\sqrt2$, where $\tan\theta_\iota$ defines the
ratio of the octet and singlet components in the $\eta(1475)$ wave
function, and $r/\sqrt2$ is the ratio of the octet and singlet
coupling constants of the $\eta(1475)$ with $\gamma\rho^0$.

Equation (5) is correct for the $\eta(1475)$ wave function of the
general form, that is, for the mixing of the $SU(3)$ octet
(quark-antiquark) and $SU(3)$ singlet (quark-antiquark and glueball)
components. This is evidenced by the presence of three parameters in
Eq. (5), $g_{\iota\gamma\rho}$, $\tan\theta_\iota$, and $r$.

The coupling constants, for which we apply the VDM and symmetries,
enter into expressions for the decay widths
$\eta(1475)\to\gamma\gamma$ and $\eta(1475)\to\gamma V$ in the
following ways:
\begin{equation}\Gamma(\eta(1475)\to\gamma\gamma)=
m^3_\iota g^2_{\iota\gamma\gamma}/64\pi\,,
\label{iotagg}\end{equation}
\begin{equation} \Gamma(\eta(1475)\to\gamma V)=
C_V[(m^2_\iota-m^2_V)/m_\iota]^3g^2_{\iota\gamma\rho}/32\pi\,
,\label{iotagV}\end{equation} where $C_\rho$\,=\,0.832 (this factor
taking into account the finite width of the $\rho^0$ resonance in
the $\eta(1475)\to\gamma\rho^0\to\gamma\pi^+\pi^-$ decay \cite{AS1};
for the stable $\rho^0$, $C_\rho$\,=\,1), $C_\omega$\,=\,1/9 and
$C_\phi$\,=\,$(2/9)H^2(x)$.

Owing to the $\eta(1475)\to\gamma\rho^0$ transition only, $\Gamma(
\eta (1475)\to\gamma\rho^0\to\gamma\gamma)\approx$\,5.9\,keV. Owing
to the $\eta(1475)\to\gamma\rho^0$ and $\eta(1475)\to\gamma\omega$
transitions, $\Gamma(\eta (1475)\to(\gamma\rho^0+\gamma\omega)\to
\gamma\gamma)\approx $\,7.3\,keV. If the $\eta(1475)$ is an $SU(3)$
singlet, then $x$\,=\,0, $H(x=0)$\,=\,1, and $\Gamma(\eta(1475)\to
(\gamma\rho^0+\gamma\omega+ \gamma\phi)\to\gamma\gamma)\approx
$\,10.5\,keV. Some restriction on $|H(x)|$ can be obtained from the
relation
\begin{eqnarray}\label{Hx}\frac{\Gamma(\eta(1475)\to\gamma\phi)}
{\Gamma(\eta(1475)\to\gamma\rho^0)}=0.1\times H^2(x)\ \,\nonumber  \\
=\xi=\frac{B(J/\psi
\to\gamma\eta(1475)\to\gamma\gamma\phi)}{B(J/\psi\to
\gamma\eta(1475)\to\gamma\gamma\rho^0)}\,. \end{eqnarray} According
to the BES experiment \cite{B1}, in which the $\gamma\rho^0$ and
$\gamma\phi$ channels have been investigated simultaneously,
$B(J/\psi\to\gamma\eta(1440)\to\gamma \gamma\rho^0)
=(1.07\pm0.17\pm0.11)\times10^{-4}$ and $B(J/\psi\to\gamma
\eta(1440)\to\gamma\gamma\phi)=(0.31 \pm0.30)\times10^{-4}$, which
corresponds to a 95\% C.L. upper limit
$B(J/\psi\to\gamma\eta(1440)\to\gamma \gamma\phi)<0.82\times10^{-4}
$ (see Table II). Hence, $\xi<0.77$ \cite{FN3}. Let, for example,
$\xi=$ 0, 0.29, 0.77. Then we get from Eq. (\ref{Hx})
$H(x)$\,=\,$-2.77,-1.7,0, 1.7, 2.77$, and from Eqs. (4)--(7),
$\Gamma(\eta(1475)\to\gamma\gamma)$\,=\,1.45, 3.2, 7.3, 13.1, 17.6
keV, respectively. (See also endnote \cite{FN4}.)

Thus, using the data on the $J/\psi\to \gamma\eta(1475)\to\gamma
\gamma(\rho^0,\phi)$ decays, we estimate
\begin{equation}
\Gamma(\eta(1475)\to\gamma\gamma)>1.45\mbox{ keV}\,,
\label{EstimateGgg}\end{equation} which is in conflict with the L3
\cite{Ac,Ach} and CLEO II \cite{Ah} results on the reaction
$\gamma\gamma\to\eta(1475)\to K\bar K\pi$ unambiguously indicative
of the suppression of the $\eta(1475)\to\gamma\gamma$ decay (see
Table I).

The above analysis allows us to conclude that the contradiction with
the results of the direct measurements of the
$\eta(1475)\to\gamma\gamma$ decay width is a real challenge.

\section{\boldmath SOLUTION OF THE $\eta(1475)\to\gamma\gamma$
DECAY PROBLEM AND EXPLANATION OF THE $\gamma\gamma^*(Q^2)\to K\bar
K\pi$ DATA}

In our early work \cite{AS2}, we showed that taking into account the
heavy vector mesons $V'$ ($V'$\,=\,$\rho'^{\,0}$,\,$
\omega'$,\,$\phi'$) in the VDM framework, along with the usual
$\rho^0$, $\omega$, and $\phi$ mesons, permits one to easily solve
the problem with $\Gamma (\eta(1475)\to\gamma\gamma)$ owing to the
strong destructive interference between the $V$ and $V'$
contributions in the $\eta(1475)\to(\gamma V+\gamma V')\to\gamma
\gamma)$ transition amplitude. Here we discuss this solution in more
detail. It is important that the proposed explanation of a number of
the experimental facts results in the nontrivial prediction
\cite{AS2}: there must arise the resonance peak caused by the
$\eta(1475)$ meson production in the $\gamma\gamma^*(Q^2)\to K\bar
K\pi$ reaction cross section for $Q^2\neq0$. Indeed, if the almost
total compensation between the $V$ and $V'$ contributions takes
place at $Q^2$\,=\,0 in $\Gamma (\eta(1475)\to\gamma\gamma
)$\,$\equiv$\,$\Gamma(\eta(1475)\to \gamma\gamma^*(Q^2=0))$, then it
is broken with increasing $Q^2$ because of the considerable $V$-$V'$
mass difference, and $\Gamma (\eta(1475)\to\gamma\gamma^*(Q^2))$
sharply increases. It is very likely that only the above phenomenon
has been observed in single-tagged two-photon interactions by the
TPC/2$\gamma$, MARK II, JADE, CELLO, CLEO II, and L3 Collaborations.

The reactions $\gamma\gamma^*(Q^2)\to K^0_SK^\pm\pi^\mp$
\cite{Be,Ac,Ah, Ach,Ai2,Gid,Ai3,Ol,Hil} have been investigated
parallel with the reactions $\gamma\gamma\to K^0_SK^\pm\pi^\mp$
\cite{Je,Al,Ai,Be,Ac,Ah,Ach}. A clear resonance signal in their
cross sections has been found in the $K^0_SK^\pm\pi^\mp$ invariant
mass range $1.35$\,GeV\,$<W<$\,1.55\,GeV for $Q^2\neq0$ (in the
region 0.04\,GeV$^2$\,$<Q^2<$\,(1--8)\,GeV$^2$) in the experiments
performed by TPC/2$\gamma$ \cite{Ai2,Ai3}, MARK II \cite{Gid}, JADE
\cite{Hil}, CELLO \cite{Be}, and CLEO II \cite{Ah}. The absence of
the resonance signal in $\sigma(\gamma\gamma\to K^0_SK^\pm\pi^\mp)$
and its appearance in $\sigma(\gamma\gamma^*(Q^2)\to K^0_SK^\pm
\pi^\mp)$ has led naturally to the hypothesis of the $J^P=1^+$
$f_1(1420)$ resonance production \cite{PDG10,Be,Ah,Ai2,Ca,Ai3,Gid,
Hil,Ol}, which is forbidden in two real photon collisions \cite{LY}.
For small $Q^2$, the $\gamma \gamma^*(Q^2)$\,$\to$\,$f_1(1420)$
transition amplitude is proportional to $\sqrt{Q^2}$
\cite{Be,Ai2,Ai3,Gid,Hil,Ol,LY,Ro,Re,Po,Ca,Sh,Ach}. However, this
can in no way eliminate the contradiction connected with the
$\eta(1475)\to\gamma\gamma,\,\gamma\rho^0,\,\rho^0\rho^0$ decays.

Poor statistics in the TPC/2$\gamma$ (12 useful events) \cite{Ai3},
MARK II (13 events) \cite{Gid}, JADE (16 events) \cite{Hil}, and
CELLO (17 events) \cite{Be} experiments did not allow determination
of the spin-parity of the resonance structure directly with the
angular distributions. The L3 \cite{Ac,Ach} and CLEO II \cite{Ah}
conclusions about the quantum numbers of the enhancement discovered
in the region 1.35--1.55 GeV are also based only on the data for the
$Q^2$ dependence of $\sigma(\gamma\gamma^*(Q^2)\to K^0_SK^\pm
\pi^\mp)$. In the last L3 experiment \cite{Ach}, $193\pm20$ events
have been selected in the resonance region. They were about evenly
distributed among five $Q^2$ intervals: $0-0.01$, $0.01-0.12$,
$0.12-0.4$, $0.4-0.9$, and $0.9-7$ GeV$^2$. The presence of events
in the first $Q^2$ interval was naturally associated with the
production of the $\eta(1475)$ \cite{FN5}, and, to describe the data
for higher $Q^2$, the contribution of the $f_1(1420)$ resonance was
recruited \cite{FN6}. As already noted above, the resonance signal
at $Q^2\approx0$ has not been observed in the CLEO II experiment
\cite{Ah}, and, therefore, the enhancement discovered for
intermediate $Q^2$ (0.04\,GeV$^2$\,$ <Q^2<$\,0.36\,GeV$^2$ and
$Q^2\gtrsim1$ GeV$^2$) was attributed without any provisos to the
$f_1(1420)$ production \cite{Ah}.

\begin{figure}\centerline{\epsfysize=3.4in
\epsfbox{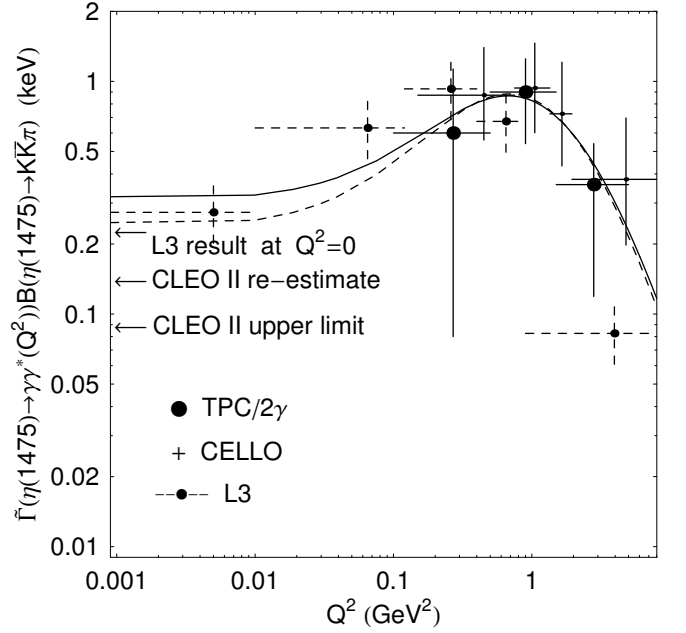}} 
\caption{$\widetilde{\Gamma}(\eta(1475)\to\gamma \gamma^*(
Q^2))B(\eta(1475)\to K\bar K\pi)$ as a function of $Q^2$. The points
with the error bars were obtained from data from TPC/2$\gamma$
\cite{Ai3}, CELLO \cite{Be}, and L3 \cite{Ach}. The arrows point to
the L3 result at $Q^2$\,=\,0 \cite{Ach} and the CLEO II upper limit
\cite{Ah}, presented in Table I, as well as the CLEO II upper limit
re-estimated by L3 \cite{Ach}, equal to 0.14 keV. The fitted curves
are described in the text.}\label{Figure1}\end{figure}

The possibility that we outlined permits one to explain the
available data on the reaction  $\gamma\gamma^*(Q^2)\to K\bar K\pi$
in the $f_1(1420)/\eta(1475)$ region by the $\eta(1475)$ resonance
production only (see Fig. \ref{Figure1}). Let us explain the
experimental points and theoretical curves shown in this figure.

The cross section for the production of a resonance $R$ with
$J^P=0^-$ or $1^+$ in $\gamma\gamma^*$ collisions can be written in
the form:
\begin{eqnarray}\label{SigmaQ2}\sigma(\gamma\gamma^*(Q^2)\to R\to K
\bar K\pi)=8\pi(2J+1)N_J\quad \nonumber \\
\times\left(1+\frac{Q^2}{m^2_R}\right)
\frac{\widetilde{\Gamma}(R\to\gamma\gamma^*(Q^2))B(R\to K\bar
K\pi)\Gamma^{tot}_R}{(m^2_R-W^2)^2+(m_R\Gamma^{tot}_R)^2}
,\end{eqnarray} where $N_0$\,=\,1, $N_1$\,=\,2, $m_R$ is the mass,
$\Gamma^{tot}_R$ the total width, and $\widetilde{\Gamma}(R\to\gamma
\gamma^*(Q^2))$ the reduced $\gamma\gamma^*$ width of the resonance.
Detailed discussions of the parametrizations and normalizations of
the $\gamma\gamma^*$ decay widths for the resonances with spin
$J$\,=\,0 and 1 may be found in Refs. \cite{Ai2,Ca,Gid,AS2,Ai3,
Be,Ach,Hil,Ol,Acha}. Necessary information is briefly presented in
the Appendix.

The data on the $Q^2$ dependence of $\widetilde{\Gamma}
(R\to\gamma\gamma^*(Q^2))$ are the matter of theoretical fits. The
TPC/2$\gamma$ \cite{Ai3}, CELLO \cite{Be}, and L3 \cite{Ach} data,
attributed to the resonance with $J^P$\,=\,$1^+$, are easily
recalculated in the case of the resonance with $J^P$\,=\,$0^-$
having approximately the same mass. (See the Appendix for details.)
The corresponding values for $\widetilde{\Gamma}
(\eta(1475)\to\gamma \gamma^*( Q^2))B(\eta(1475)\to K\bar K\pi)$ are
shown in Fig. \ref{Figure1} \cite{FN7}.

To describe $\widetilde{\Gamma}(\eta(1475)\to\gamma\gamma^*(Q^2))
B(\eta(1475)\to K\bar K\pi)$, we use the following parametrization:
\begin{eqnarray}\label{GammaQ2}\widetilde{\Gamma}(\eta(1475)\to\gamma
\gamma^*(Q^2))B(\eta(1475)\to K\bar K\pi)\ \nonumber \\
=\left|\,A\left(\frac{10}{9}\,\frac{1}{1+Q^2/m^2_\rho}+
\frac{2}{9}\,\frac{h}{1+Q^2/m^2_\phi}\right)\right.\ \ \ \nonumber  \\
\left.\ \ \ +A'\left(\frac{10}{9}\,\frac{1}{1+Q^2/m^2_{\rho'}}+
\frac{2}{9}\,\frac{h}{1+Q^2/m^2_{\phi'}}\right)\right|^2,
\end{eqnarray} where $m_{\rho'}$\,=\,1.45 GeV \cite{PDG10},
$m_{\phi'}$\,=\,1.68 GeV \cite{PDG10}, a parameter $h$\,=\,$H(x)$
[see Eq. (\ref{VDM1})], $$A^2=\Gamma(\eta(1475)\to\gamma
\rho^0\to\gamma\gamma)B(\eta(1475)\to K\bar K\pi),$$ and
$$A'=-A-9\frac{\sqrt{\Gamma(\eta(1475)\to\gamma\gamma)B(\eta
(1475)\to K\bar K\pi)}}{(10+2h)};$$
$\Gamma(\eta(1475)$\,$\to$\,$\gamma\gamma)=\widetilde{\Gamma}
(\eta(1475)$\,$\to$\,$\gamma\gamma^*(Q^2=0))$. In Eq.
(\ref{GammaQ2}), we assume for simplicity that the $V'$ meson family
has the ideal nonet structure, too. If we fix the value of
$\Gamma(\eta(1475)$\,$\to$\,$\gamma\gamma)B(\eta(1475)$\,$\to$\,$K\bar
K\pi)$, then Eq. (\ref{GammaQ2}) will contain only two free
parameters $A$ and $h$. For example, if we put (according to L3
\cite{Ach}) $\Gamma(\eta(1475)$\,$\to$\,$\gamma\gamma)
B(\eta(1475)$\,$\to$\,$K\bar K\pi)$\,=\,0.23 keV, then the fitting
to the data gives $A$\,=\,2.44 keV$^{1/2}$ and $h$\,=\,$-1.7$
($\chi^2$\,=\,5 for 9 degrees of freedom; see the Appendix), and, at
$r=1$, $\theta_\iota\approx85.5^\circ$. The result of the fit is
shown in Fig. \ref{Figure1} by the dashed curve. It is interesting
to note that $h$\,=\,$-1.7$ from the fit is in agreement with the
estimate $|H(x)|<2.77$ obtained above from the BES data \cite{B1}
and that the value of $A^2$, at $B(\eta(1475)$\,$\to$\,$K\bar
K\pi)\approx1$, corresponds to $\Gamma (\eta(1475)$\,$\to$\,$\gamma
\rho^0)\approx1$ MeV, i.e., it is in agreement with the estimate in
Eq. (\ref{EstimateGigr}). If $\Gamma(\eta(1475)$\,$\to
$\,$\gamma\gamma)B(\eta(1475) $\,$\to$\,$K\bar K\pi)$ is not fixed,
i.e., if $A'$ is considered as a free parameter, then the fit, shown
in Fig. \ref{Figure1} by the solid curve, gives $A$\,=\,1.88
keV$^{1/2}$, $A'$\,=\,2.48 keV$^{1/2}$, and $h$\,=\,$-0.9$
($\Gamma(\eta(1475)$\,$\to$\,$\gamma \gamma)B(\eta(1475)$\,$\to$\,$
K\bar K\pi)$\,=\,0.3 keV, $\chi^2$\,=\,3.9 for 8 degrees of
freedom); and at $r=1$, $\theta_\iota\approx68^\circ$. In this case,
the value of $A^2$ corresponds to $\Gamma(\eta(1475)\to\gamma\rho^0)
\approx1$ MeV, if $B(\eta(1475)$\,$\to$\,$K\bar K\pi)\approx0.6$.

\begin{figure}\centerline{\epsfysize=3.4in
\epsfbox{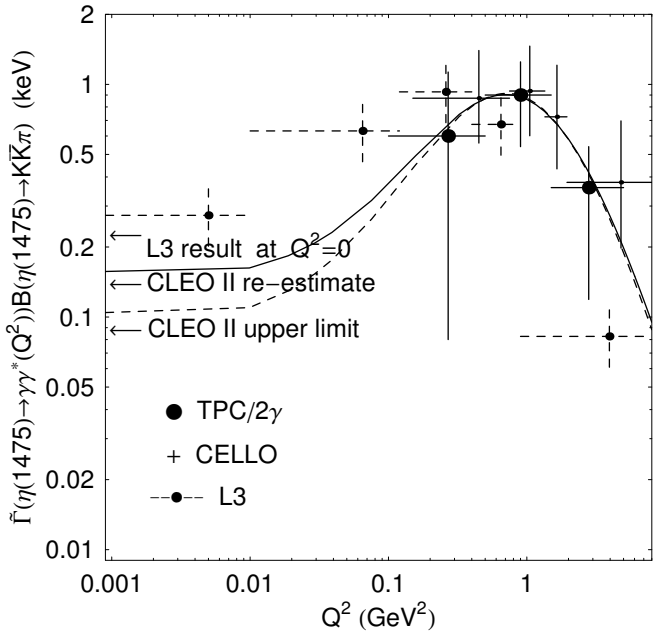}} 
\caption{The same as in Fig. \ref{Figure1} but with different
fitting variants as described in the
text.}\label{Figure2}\end{figure}

Let us consider two more variants different in normalization at
$Q^2$\,=\,0. Take $\Gamma(\eta(1475)$\,$\to$\,$\gamma\gamma
)B(\eta(1475)$\,$\to$\,$K\bar K\pi)$\,=\,0.14 keV or =\,0.089 keV,
which is equal to the CLEO II upper limit re-estimated by L3
\cite{Ach} (see Sec. II) or the CLEO II upper limit \cite{Ah} (see
Table I), respectively. The fits obtained for these variants are
shown in Fig. \ref{Figure2} by the solid curve ($A$\,=\,3.22
keV$^{1/2}$, $h$\,=\,$-2.36$; at $r=1$,
$\theta_\iota\approx-86^\circ$) and the dashed curve ($A$\,=\,3.75
keV$^{1/2}$, $h$\,=\,$-2.66$; at $r=1$, $\theta_\iota\approx
-83^\circ$), respectively. For $\Gamma
(\eta(1475)$\,$\to$\,$\gamma\gamma)B(\eta(1475)$\,$\to$\,$ K\bar
K\pi)$\,=\,0.14 keV, the L3 points in the region $Q^2<0.1$ GeV$^2$
are described within 2 standard deviations. The description of the
higher $Q^2$ region is quite satisfactory for both normalizations.

When accurate data on the $\eta(1475)$ in $\gamma\gamma^*$
collisions appears, the model for the $Q^2$ dependence of
$\widetilde{\Gamma}(\eta(1475)\to\gamma\gamma^*(Q^2))$ can be
refined. For example, the masses $m_{\rho'}$ and $m_{\phi'}$ could
be considered as free parameters, and their values might be defined
from a fit. Theoretically, high-quality data can allow us to
consider even heavier vector meson contributions.

Thus, with the help of the $\eta(1475)$, one can explain
simultaneously the peak in the $\gamma\rho^0$ mass spectrum in the
$J/\psi\to\gamma\gamma\rho^0$ decay, the pseudoscalar structures
near thresholds in the $\rho\rho$ and $\omega\omega$ mass spectra in
$J/\psi\to\gamma(\rho\rho,\omega\omega)$ decays \cite{AS1}, and the
suppression of the $\eta(1475)$ signal in the reaction $\gamma\gamma
\to K\bar K\pi$ and its appearance in $\gamma\gamma^*(Q^2)\to K\bar
K\pi$ for $Q^2\neq0$. Our explanation might be rejected
unambiguously by measuring the spin-parity of the signal in the
region of 1475 MeV in the reaction $\gamma\gamma^*(Q^2)\to K\bar
K\pi$, together with the disavowal of the pseudoscalar structures in
the $\rho\rho$ and $\gamma\rho^0$ mass spectra in the
$J/\psi\to\gamma \rho\rho$ and $J/\psi\to\gamma\gamma\rho^0$ decays.

\section{\boldmath $f_1(1420)\to\gamma(\rho^0,\phi)$ AND $\gamma\gamma^*$ DECAYS}

Here we discuss the $f_1(1420)$ production in the reaction
$\gamma\gamma^*(Q^2)\to f_1(1420)\to K\bar K\pi$ using the
information about $f_1(1420)\to\gamma(\rho^0,\phi)$ decays as a
guide.

We note, for short, $f_1(1420)$ ($f_1(1285))$ by $f'_1$ ($f_1$) and
write the $f'_1\to\gamma V$ decay width in the form
\begin{eqnarray}\label{f1-gV} \Gamma(f'_1\to\gamma
V)=C^{f'_1}_V\,\widetilde{\Gamma}(f'_1\to\gamma V)\ \ \nonumber \\
\times\frac{m^2_V}{m^2_{f'_1}}\left(1-\frac{m^2_V}
{m^2_{f'_1}}\right)^3 \left(1+\frac{m^2_V}{m^2_{f'_1}}\right),
\end{eqnarray} where $C^{f'_1}_\rho$\,=\,0.74 takes into account the
finite width of the $\rho^0$ resonance in the $f'_1
\to\gamma\rho^0\to\gamma\pi^+\pi^-$ decay, $C^{f'_1
}_\omega$\,=\,$C^{f'_1 }_\phi$\,=\,1, and the last factor includes
the contributions from the longitudinal and transverse $V$ meson
production, respectively, as in the model \cite{Ca}. Because the
quantities $\widetilde{\Gamma}(f'_1\to\gamma V)$ are free from the
main kinematical factors, we shall assume that they are independent
of $m^2_V$. Applying to them the na\"{\i}ve quark model, we have
\begin{equation}\label{qq-f1-gV}
\widetilde{\Gamma}(f'_1\to\gamma\rho^0)=\frac{9}{4}\,
\widetilde{\Gamma}(f'_1\to\gamma\phi)\tan^2(\theta_i-\theta_A)
\end{equation} and $\widetilde{\Gamma}(f'_1\to\gamma\omega)=
\widetilde{\Gamma}(f'_1\to\gamma\rho^0)/9$, where
$\theta_i$\,=\,$35.3^\circ$ is the ``ideal'' mixing angle and
$\theta_A$ is the mixing angle in the axial-vector nonet, the
members of which are $f'_1$ and $f_1$. To estimate
$B(f'_1\to\gamma\rho^0)$, we use Eqs. (\ref{f1-gV}),
(\ref{qq-f1-gV}), $B(f'_1\to\gamma\phi)/B(f'_1\to K\bar K\pi)$\,=\,$
0.003\pm0.001\pm0.001$ \cite{Ba,FN8}, and $\Gamma^{tot}_{f'_1}
$\,=\,$(54.9\pm2.6)$ MeV \cite{PDG10}. This gives
$\Gamma(f'_1\to\gamma\phi )\approx(0.16\pm0.08)B(f'_1\to K\bar
K\pi)$ MeV. The mixing angle $\theta_A$ can be found in a variety of
ways, that is, from the different mass formulae as well as from
suitable data (if they exist) on decays and production reactions.
Every way is accompanied by some specific assumptions. For example,
the Gell-Mann-Okubo-Sakurai mass formula \cite{PDG10,GMOS}
\begin{equation}\label{GMOS} m^2_{f'_1}\cos^2\theta_A+m^2_{f_1}\sin^2
\theta_A=\frac{1}{3}(4m^2_{K_{1A}}-m^2_{a_1})\end{equation} gives
$\theta_A$\,=\,$37.9^\circ\pm5^\circ$, where the quark model
considerations concerning the sign of $\theta_A$ \cite{PDG10} and
the mass values from the PDG review \cite{PDG10} have been used. The
error in $\theta_A$ has been estimated from the $a_1$ mass
uncertainty. On the other hand, the model, in which the
octet-singlet mixing is due to the symmetry breaking for the mass
particles \cite{PDG10,Sch}, gives
\begin{equation}
\tan\theta_A=\frac{4m^2_{K_{1A}}-m^2_{a_1}-3m^2_{f'_1}}{2\sqrt{2}
(m^2_{a_1}-m^2_{K_{1A}})}\end{equation} and
$\theta_A=29.2^\circ\pm5.5^\circ$. Then, from Eqs. (\ref{f1-gV}) and
(\ref{qq-f1-gV}) for the central values of
$B(f'_1\to\gamma\phi)/B(f'_1\to K \bar K\pi)$\,=\,0.003 and
$\theta_A$\,=\,$37.9^\circ$ ($29.2^\circ$), we get
$B(f'_1\to\gamma\rho^0)/B(f'_1\to K \bar K\pi)=1.6\times10^{-5}$
($8.4\times10^{-5}$) and, for $B(f'_1\to\gamma\phi)/B(f'_1\to K \bar
K\pi)$\,=\,$0.003+0.0014$, $\theta_A=29.2^\circ-5.5^\circ$,
\begin{equation}\label{f1grho} B(f'_1\to\gamma\rho^0)/B(f'_1\to K
\bar K\pi)=4.6\times10^{-4}.\end{equation} It is the maximal
increase of the ratio within one error in every factor. This
estimate should be compared with the available experimental
restriction $B(f'_1\to\gamma\rho^0)/B(f'_1\to K \bar K\pi)<0.02$
\cite{Ba}, which was discussed in Sec. III.

We now estimate the quantity $\widetilde{\Gamma}(f'_1\to\gamma
\gamma)$ (see Eq. (\ref{W1gg}) in the Appendix), due to the
$f'_1\to\gamma V$ transitions. (Note that, in contrast to the
$\eta(1475)$, there has been no constructive idea that speculates
about heavier vector meson contributions in the $f'_1\to\gamma
\gamma^*$ decay for a while.) According to the VDM,
\begin{eqnarray}\label{f1-gV-gg}
\widetilde{\Gamma}(f'_1\to\gamma\gamma)\qquad\qquad\qquad\qquad
\nonumber \\ =\frac{4\pi
\alpha}{9f^2_\rho}\,\widetilde{\Gamma}(f'_1\to\gamma\phi)
\left(1-\frac{5}{\sqrt{2}}\tan(\theta_i-\theta_A)\right)^2, &&
\end{eqnarray} and, for the central value of
$B(f'_1\to\gamma\phi)/B(f'_1\to K \bar K\pi)$\,=\,0.003,
\begin{equation}\label{f1ggEstim} \widetilde{\Gamma}(f'_1\to\gamma\gamma)
\approx(1,\, 0.77,\, 0.3\mbox{ keV})B(f'_1\to K \bar K\pi),
\end{equation} for $\theta_A$\,=\,$37.9^\circ$,
$35.3^\circ(=\theta_i)$, $29.2^\circ$, respectively.

Experimental results for $\widetilde{\Gamma}(f'_1\to \gamma\gamma
)B(f'_1\to K \bar K\pi)$ are presented in Table III. The values of
$\widetilde{\Gamma}(f'_1\to \gamma\gamma)B(f'_1\to K \bar K\pi)$
found from the fits have the order of 1 keV. As can be seen from
Table III, the values reveal a high sensitivity to the form assumed
for the form factor $F(Q^2)$, although $\widetilde{\Gamma}(f'_1\to
\gamma\gamma)B(f'_1\to K \bar K\pi)$ is independent of $F(Q^2)$ by
definition. [See Eq. (\ref{W1gg}) in the Appendix.] Such a situation
is due to limited statistics for the reaction
$\gamma\gamma^*(Q^2)\to f_1(1420)\to K\bar K\pi$ at low $Q^2$. Note
that the maximal value for $\widetilde{\Gamma} (f'_1\to\gamma
\gamma)B(f'_1\to K \bar K\pi)$\,=\,$(3.2\pm0.6 \pm0.7)$ keV has been
found in the most statistically significant experiment performed by
the L3 Collaboration (see Table III).

\begin{table} 
\caption{The results for $\widetilde{\Gamma}(f'_1\to\gamma\gamma)
B(f'_1\to K \bar K\pi)$. (The TPC/2$\gamma$ convention \cite{Ai3,Ol}
is used for normalization.) The results are dependent on the form
factor, shown in the fourth column: $F(Q^2)$:
$F(Q^2)$\,=\,$F_\rho$\,=\,$1/(1+Q^2/m^2_\rho)$, or
$F(Q^2)$\,=\,$F_\phi$\,=\,$1/(1+Q^2/m^2_\phi)$, or
$F(Q^2)$\,=\,$F_{L3}$\,=\,$1/[(1+Q^2/(0.926\,\mbox{GeV})^2)^2
(1+Q^2/m^2_{f'_1})^{1/2}]$.}\vspace{0.05cm}
\begin{tabular}{|c|c|c|c|} \hline \vspace{-0.35cm} & & & \\ Experiment &  Ref. &
$\widetilde{\Gamma}(f'_1\to\gamma\gamma) \ $ & $F(Q^2)$
\\ &  & $\ \times B(f'_1\to K\bar K\pi)$ (keV)\ & \\ \hline
MARK II,       1987 & \cite{Gid}  &  $1.6\pm0.7\pm0.3$           & $F_\rho$ \\
&             &  $1.1\pm0.5\pm0.2$           & $F_\phi$  \\ \hline
TPC/2$\gamma$, 1988 & \cite{Ai3}  &  $1.3\pm0.5\pm0.3$           & $F_\rho$ \\
&             &  $0.63\pm0.24\pm0.15$        & $F_\phi$  \\ \hline
JADE,          1989 & \cite{Hil}  &  $2.3\pm^{1.0}_{0.9}\pm0.8$  & $F_\rho$ \\
&             &  $1.5\pm^{0.6}_{0.5}\pm0.5$  & $F_\phi$  \\ \hline
CELLO,         1989 & \cite{Be}   &  $1.5\pm0.5\pm0.4$           & $F_\rho$ \\
&             &  $0.7\pm0.2\pm0.2$           & $F_\phi$  \\ \hline
L3,            2007 & \cite{Ach}  &  $3.2\pm0.6\pm0.7$           & $F_{L3}$ \\
\hline \end{tabular}\end{table}

A comparison of our tentative estimate (\ref{f1ggEstim}) with highly
model-dependent data from Table III does not allow conclusions about
the $f_1(1420)$ dominance in the reaction $\gamma \gamma^*(Q^2)\to
K\bar K\pi$. It is impossible to exclude that there are two
resonance contributions in this reaction, from the pseudoscalar
$\eta(1475)$ and the axial-vector $f_1(1420)$, whose separation
requires substantial improvement of the data and obligatory
determination of the spin-parity.

\vspace{-0.1cm}
\section{CONCLUSION AND OUTLOOK}

We resolved the contradiction between the suppression of the
$\eta(1475)\to\gamma\gamma$ decay and the strong couplings of the
$\eta(1475)$ to the $\rho\rho$, $\omega\omega$, and $\gamma\rho^0$
channels by taking into account the effect of the heavy vector
mesons $\rho'^{\,0}$, $\omega'$, $\phi'$. This led us to the
explanation of the resonancelike $Q^2$ dependence on the
$\gamma\gamma^*(Q^2)\to K\bar K\pi$ reaction cross section by
$\eta(1475)$ production, which is an alternative to the conventional
explanation of dependence by $f_1(1420)$ production.

To check our scenario and resolve the difficulties accumulated in
understanding properties of the $\eta(1475)$, further experimental
investigations are required:

\begin{description}
\item{(1)} measurements of spin-parities of the intermediate states in
the reaction $\gamma\gamma^*(Q^2)\to K\bar K\pi$ in the $\eta(1475)$
region for $0\lesssim Q^2\lesssim3$ GeV$^2$ (which implies the
separation of pseudoscalar and pseudovector contributions by using
the angular distributions),
\item{(2)} further high-statistics measurements of the pseudoscalar
structures in the $\rho\rho$ and $\omega\omega$ mass spectra near
their thresholds in the $J/\psi\to\gamma\rho\rho$ and $J/\psi\to
\gamma\omega\omega$ decays,
\item{(3)} a reliable determination of the spin of the $\gamma\rho^0$ system
in the $J/\psi\to\gamma R\to\gamma\gamma\rho^0$ decay in the region
of 1.475 GeV,
\item{(4)} acquisition accurate data on the $\eta(1475)/f_1(1420)\to\gamma
\phi$ decays.
\end{description}

High-statistics experiments necessary to solve these problems seem
feasible at $B$ and $C/\tau$ factories with the Belle, $BABAR$, CLEO
II, and BES III detectors. \vspace{0.1cm}

$$\mbox{\small \bf ACKNOWLEDGMENTS}$$

This work was supported in part by RFFI Grant No. 10-02-00016 from
the Russian Foundation for Basic Research. \vspace{-0.1cm}
$$\mbox{\small \bf APPENDIX}$$

\setcounter{equation}{0}
\renewcommand{\theequation}{A\arabic{equation}} 

The formation of a $0^-$ state proceeds in collisions of two
transverse photons. The $0^-\to\gamma\gamma^*$ decay width is
\begin{eqnarray}\label{W0Q2gg}\Gamma(0^-\to\gamma\gamma^*(Q^2))
\qquad\qquad\ \nonumber \\
=(1+Q^2/m^2_{0^-})^3\widetilde{\Gamma}(0^-\to\gamma\gamma^*(
Q^2)).\end{eqnarray} Here the factor $(1+Q^2/m^2_{0^-})$ comes from
the $\gamma\gamma^*$ phase space and the factor
$(1+Q^2/m^2_{0^-})^2$ comes from the gauge-invariant Lorentz
structure $\varepsilon_{\mu\nu\sigma\tau}q_\mu
\epsilon_\nu(q)q^*_\sigma \epsilon_\tau(q^*)$ associated with the
$0^-\gamma\gamma^*$ vertex, where $\epsilon_\nu(q)$ and
$\epsilon_\tau(q^*)$ are the polarization four-vectors of the
photons $\gamma$ and $\gamma^*$ with four-momenta $q$ and $q^*$,
respectively; $Q^2$\,=\,$-q^{*2}$.

The $1^+$ state can be produced either through collision of one
longitudinal and one transverse (LT) or through two transverse (TT)
photons. Thus, in the general case, the $1^+\to\gamma\gamma^*$ decay
width is expressed in the terms of two independent functions of
$Q^2$:
\begin{eqnarray}\label{W1Q2gg}\Gamma(1^+\to\gamma\gamma^*( Q^2))
\qquad\qquad\quad\nonumber \\ =(1+Q^2/m^2_{1^+})^3\left[
\widetilde{\Gamma}^{\mbox{\scriptsize{LT}}}(Q^2)+
\widetilde{\Gamma}^{\mbox{\scriptsize{TT}}}(Q^2)\right],\end{eqnarray}
where the factor $(1+Q^2/m^2_{1^+})^3$ \cite{Be,Ca,Ol,Ro} has its
origin similar to the $0^-$ case. The amplitude for the
$1^+\to\gamma\gamma^*$ transition can be written as
\begin{eqnarray}\label{M1gg}M=Q^2\varepsilon_{\mu\nu\sigma\tau}
q_\mu\epsilon_\nu(q)\epsilon_\tau(q^*)\left[\xi_\sigma(p)G_1(Q^2)\right.
\nonumber \\ \left.+\,q^*_\sigma\,G_2(Q^2)\,(\xi(p),q-q^*)\right],
\qquad\quad\end{eqnarray} where $\xi(p)$ is the polarization
four-vectors of the $1^+$ resonance state, $p$\,=\,$q+q^*$,
$G_1(Q^2)$ and $G_2(Q^2)$ are the invariant functions. For
$Q^2\to0$, $\widetilde{\Gamma}^{\mbox{\scriptsize{LT}}}( Q^2)\sim
Q^2$ and $\widetilde{\Gamma}^{\mbox{\scriptsize{TT}}}(Q^2)\sim Q^4$.
The measured cross section [see Eq. (\ref{SigmaQ2})] is defined by
the combination $\widetilde{\Gamma}(1^+\to\gamma\gamma^*(Q^2))
$\,=\,$ \widetilde{\Gamma}^{\mbox{\scriptsize{LT}}}(Q^2)+
\frac{1}{2}\widetilde{\Gamma}^{\mbox{\scriptsize{TT}}}(Q^2)$, which
is usually parametrized in the following way:
\begin{eqnarray}\label{W1gg}\widetilde{\Gamma}(1^+\to\gamma\gamma^*
(Q^2))=\frac{Q^2}{m^2_{1^+}}\widetilde{\Gamma}(1^+\to\gamma\gamma)
\nonumber\\ \times\left[1+\frac{Q^2}{2m^2_{1^+}}\right]F^2(Q^2),
\end{eqnarray}\\[0.1cm] where $\widetilde{\Gamma}
(1^+\to\gamma\gamma)$ is a parameter characterizing the strength of
the coupling of the $1^+$ resonance to the $\gamma\gamma$ system,
and $F(Q^2)$ is some model form factor satisfying the normalization
condition $F(0)$\,=\,1; specific examples for $F(Q^2)$ are given in
Table III. Such a representation corresponds to the model \cite{Ca},
in which $G_2(Q^2)$\,=\,0 and
$\widetilde{\Gamma}^{\mbox{\scriptsize{TT}}}( Q^2)=(Q^2/ m^2_{1^+}
)\widetilde{\Gamma}^{\mbox{\scriptsize{LT}}}(Q^2)$.

As we mentioned in Sec. IV, the TPC/2$\gamma$ \cite{Ai3} and CELLO
\cite{Be} data, attributed to the resonance with $J^P$\,=\,$1^+$,
are easily recalculated in the case of the resonance with
$J^P$\,=\,$0^-$ having approximately the same mass. To do this, the
TPC/2$\gamma$ data (see Fig. 10(b) in the second paper from Ref.
\cite{Ai3}) should be multiplied by 6, and the CELLO data (see Fig.
8 from Ref. \cite{Be}), taking into account a difference in
normalization, should be multiplied by $3/(1+Q^2/m^2_\iota)^3$. As
for the L3 data \cite{Ach}, they have been presented for the $Q^2$
dependence of the number of $e^+e^-\to e^+e^-K^0_SK^\pm\pi^\mp$
events in the peak observed in the $f_1(1420)/\eta(1475)$ region.
The two-photon width, or two-photon coupling parameter,
$\Gamma_{\gamma\gamma}$, times the branching ratio for the $K\bar
K\pi$ decay, $BR(K\bar K\pi)$ (in the notations of Ref. \cite{Ach}),
have been extracted from the fit for the $0^-$ and $1^+$ resonances,
together with parameters of the form factors $F^2_{0^-}(Q^2)$ and
$F^2_{1^+}(Q^2)$ \cite{FN5,FN6}. Let us use this information and
make up the combination $$([\Gamma_{\gamma \gamma}BR(K\bar
K\pi)]_{0^-}F^2_{0^-}(Q^2)\qquad\qquad\qquad\qquad$$
\hspace{0.7cm}$+6[\Gamma_{\gamma \gamma}BR(K\bar
K\pi)]_{1^+}F^2_{1^+}(Q^2))/(1+Q^2/ m^2_{0^-}).$\\[0.3cm] (We ignore a
small mass difference of the $0^-$ and $1^+$ contributions.)
Averaging this combination over the five $Q^2$ intervals considered
in Ref. \cite{Ach} (and listed in Sec. IV) and taking into account
its $\approx$\,30\% error, we get putative ``experimental points''
from L3 for $\widetilde{\Gamma}(\eta(1475) \to\gamma\gamma^*(Q^2))
B(\eta(1475)\to K\bar K\pi)$. Notice that the log-scale for $Q^2$ is
used in Fig. \ref{Figure1} to emphasize the small and moderate $Q^2$
regions. The rightmost point from L3 ($0.082\pm0.024$ keV) relating
to the $Q^2$ interval $0.9-7$ GeV$^2$ is inconsistent with the other
data. An average fitted to the data is equal to $0.52\pm0.10$ keV,
(see Fig. \ref{Figure1}), and we exclude this point from our
treatment. It is clear that the available data require further
refinements.



\begin{thebibliography}{99}
\bibitem{FN1} It is sometimes suggested that $\eta(1475)$ is the
$s\bar s$ partner of the $(u\bar u+d\bar d)/\sqrt{2}$ $\eta(1295)$
state; see \cite{MCU} for review. Then, using the L3 data
\cite{Ach}, we estimate $$\Gamma(\eta(1295)\to\gamma \gamma)\qquad
\qquad$$ $$=\frac{25}{2}\left(\frac{M(\eta(1295))}{M(\eta(1475))}
\right)^3\Gamma(\eta(1475)\to\gamma\gamma)$$ $$\geq\,(1.9\pm0.6)
\mbox{ keV.}\qquad\qquad$$ However, as yet the $\eta(1295)$\,$\to
$\,$\gamma\gamma$ decay has not been observed: $\Gamma(\eta
(1295)$\,$\to $\,$\gamma\gamma)[B(\eta (1295)\to\eta\pi\pi
)+B(\eta(1295)\to K\bar K\pi)]<0.08$ keV \cite{Ac,Ah,PDG10}.
\bibitem{FN2} The $\eta(1475/1440)\to K\bar K\pi$ decay is dominant
\cite{PDG10,MCU,B3}. For example, according to the BES data
\cite{B3}, its fraction exceeds 80\%. Putting $B(\eta(1475)\to K\bar
K\pi)=0.6 $, we certainly do not overestimate
$\Gamma(\eta(1475)\to\gamma \rho^0)$.
\bibitem{FN3} A very weak experimental restriction for $B(J/\psi\to\gamma
\eta(1440)\to\gamma\gamma\omega)$ (see Table II) does not yield any
additional information.
\bibitem{FN4} For fixed $|H(x)|$, the width
$\Gamma (\eta(1475)\to\gamma\gamma)$ possesses the minimal value for
$H(x)<0$; see Eqs. (\ref{VDM1}) and (\ref{iotagg}). Let us consider
the simplest variant of the quark model in which the parameter
$x=r\tan\theta_\iota/\sqrt2=\tan\theta_\iota/\sqrt2$, i.e., $r=1$,
which corresponds to the nonet symmetry assumption for the
$\eta(1475)$ meson interaction. Then, with $H(x)$\,=\,0, we have
$x=1/2$, $\tan\theta_\iota=1/\sqrt2$, and the mixing angle
$\theta_\iota =35.3^\circ$, i.e., equals to the so-called ideal
mixing angle. If $H(x)$\,=\,$-1.7$ or $-2.77$, then $x=9$,
$\tan\theta_\iota\approx 12.7$, $\theta_\iota\approx85. 5^\circ$ or
$x\approx-4.9$, $\tan\theta_\iota \approx-6.9$,
$\theta_\iota\approx-82^\circ$, respectively. The octet component of
the $\eta(1475)$ wave function is dominant in the two cases.
\bibitem{FN5} It was postulated in Ref. \cite{Ach} that the
$Q^2$ dependence of the $\eta(1475)$ production cross section is
defined by the form factor $F^2_{0^-}(Q^2)$\,=\,$1/(1+Q^2/(1.47
\,\mbox{GeV})^2)^2$.
\bibitem{FN6} The form factor $F^2_{1^+}(Q^2)$\,=\,$(Q^2/m^2_{ 1^+})
(1+Q^2/2m^2_{1^+})/(1+Q^2/(0.926\mbox{\,GeV})^2)^4$ was used in Ref.
\cite{Ach} to fit the $Q^2$ dependence of the $f_1(1420)$ production
cross section.
\bibitem{FN7} Notice that a single resonance peak has been observed
in the $f_1(1420)/\eta(1475)$ region in all experiments on the
reaction $\gamma\gamma^*\to K\bar K\pi$ \cite{Ai3,Be,Hil,Ac,Ach,
Ah}. Certainly, one must keep in mind poor statistics and wide steps
in $W$ in most experiments that made it impossible to resolve the
overlapping peaks from the $0^-$ and $1^+$ resonances in
$\gamma\gamma^*\to K\bar K\pi$ (if they are produced simultaneously)
owing to their small mass difference.
\bibitem{FN8} We combine the statistical and systematic errors
quadratically.
\end{thebibliography}
\end{document}